\documentclass[aps,prl,twocolumn,superscriptaddress]{revtex4}

\usepackage{makeidx}
\usepackage{amsmath}
\usepackage{amsthm}
\usepackage{amssymb}
\usepackage{amsxtra}
\usepackage{amscd}
\usepackage[mathscr]{eucal}
\usepackage{boxedminipage}
\usepackage{bbm}
\usepackage{epsfig}
\usepackage{epic,eepic}
\usepackage{latexsym}
\usepackage{graphics}
\newcommand{\bra}[1]{\ensuremath{\langle#1|}}
\newcommand{\ket}[1]{\ensuremath{|#1\rangle}}
\newcommand{\be}{\begin{equation}}
\newcommand{\ee}{\end{equation}}
\newcommand{\DBCS}{\ensuremath{\Delta_\mathrm{BCS}}}

\newcommand{\oD}{\ensuremath{\omega_\mathrm{co}}}

\newcommand{\olcite}[1]{[\onlinecite{#1}]}
\newcommand{\wL}{\ensuremath{w_\mathrm{L}}}

\newcommand{\Eq}[1]{Eq.~(\ref{#1})}
\newcommand{\eq}[1]{Eq.~(\ref{#1})}
\newcommand{\brags}{\ensuremath{\langle \mathrm{gs}|}}
\newcommand{\ketgs}{\ensuremath{|\mathrm{gs}\rangle}}
\newcommand{\Fig}[1]{Fig.~\ref{#1}}

\newcommand{\pdag}{{\phantom{\dagger}}}

\begin{document}

\title{Well-defined quasiparticles in interacting metallic grains}

\author{Dominique Gobert}
\email{gobert@lmu.de}
\affiliation{Physics Department and CeNS, LMU M\"unchen,
  Theresienstr.\ 37, 80333 M\"unchen, Germany}
\affiliation{Institute for Theoretical Physics C, RWTH Aachen, 52056 Aachen, Germany}
\author{Moshe Schechter}
\affiliation{Department of Physics and Astronomy, University of
  British Columbia, Vancouver, British Columbia, Canada V6T 1Z1}
\author{Ulrich Schollw\"ock}
\affiliation{Institute for Theoretical Physics C, RWTH Aachen, 52056 Aachen, Germany}
%\email{scholl@theorie.physik.uni-muenchen.de}
\author{Jan von Delft}
\affiliation{Physics Department and CeNS, LMU M\"unchen, Theresienstr.\ 37, 80333 M\"unchen, Germany}
%\email{vondelft@theorie.physik.uni-muenchen.de}
%\homepage[]{Your web page}
%\thanks{}
\date{\today}

\begin{abstract}
  
  We analyze spectral functions of mesoscopic systems with large
  dimensionless conductance, which can be described by a universal
  Hamiltonian.
  We show that an important class of spectral functions are dominated by one
  single state only, which implies the existence of well-defined
  (i.e.~infinite-lifetime) quasiparticles.  
  Furthermore, the dominance of a single state enables us to calculate
  zero-temperature spectral functions with high accuracy using the
  density-matrix renormalization group.  
  We illustrate the use of this
  method by calculating the tunneling density of states of metallic
  grains and the magnetic response of mesoscopic rings.
%, of which we discuss
%  the crossover from the few-electron to the bulk regime.
\end{abstract}

%check pacs!
% PACS, the Physics and Astronomy
                             % Classification Scheme.

\maketitle

The pairing Hamiltonian of Bardeen, Cooper and Schrieffer (BCS) is
established as the paradigmatic framework for describing
superconductivity \cite{CooperSchrieffer57, Tinkham96}.  
The BCS {\emph{solution}} is, however, an approximate one, valid (and
exceedingly successful) only as long as the mean level spacing $d$ is
much smaller than the superconducting band gap $\DBCS$
\cite{DelftRalph01, SchechterDelft01}.  
One of the main features of
the BCS solution is the description of the excitation spectrum by
well-defined (i.e.~infinite-lifetime) Bogoliubov quasiparticles,
responsible for many of the features of the superconducting state.

In this Letter, we address the question whether this quasiparticle 
picture prevails in the
entire regime of parameters -- including the case that the samples are so
small or so weakly interacting that the BCS solution is inapplicable
-- by analyzing spectral functions.  
For example, the spectral
function corresponding to the (noninteracting) particle creation
operator $c^\dagger_{k \sigma}$ is 
given, within the BCS solution, by a sharp line in $k$-$\omega$-space;
this reflects the infinite lifetime of the quasiparticles.
For an interacting system, this is a very peculiar property, since the
interactions usually shift a significant portion of the spectral
weight to a background of excitations, responsible for the finite lifetime
of the quasiparticles.
Here we show that the unusual property of finding well-defined 
quasiparticles
persists to a very good approximation over the entire parameter range
of the pairing Hamiltonian, and is not merely a property of the mean
field approximation in the BCS regime.
We also give a condition for more general spectral functions to show
analogous behaviour.

Of central importance is that this 
result is relevant not only in the context of mesoscopic
superconductivity, but more generally for disordered systems with
large dimensionless conductance $g$ (defined as the ratio between the
Thouless energy and the mean level spacing $d$).
This is because to lowest order in $g^{-1}$,
the electron-electron interactions can be described by a remarkably simple 
universal Hamiltonian 
(UH) 
\cite{KurlandAltshuler00, AleinerGlazman02},
which has, besides the kinetic energy term
$H^0 = \sum_{i \sigma} \epsilon_i c^\dagger_{i\sigma} c^\pdag_{i\sigma}$,
only three couplings:
\begin{eqnarray}
\label{pairing_Ham}
H^c \!= \!E_c \hat{n}^2, \; H^s \!= \!J_s \hat{\vec{S}}^2, \;
 H^p \!=   \!- \lambda d \!\! \sum_{i, j \in \cal{N}}  
           c^\dagger_{i\uparrow} c^\dagger_{i\downarrow}
           c^\pdag_{j\downarrow} c^\pdag_{j\uparrow}.
\end{eqnarray}
Here, $E_c$, $J_s$ and $-\lambda d$ are coupling constants.
%Although $\lambda$ may have either sign, our analysis is restricted to
%the case of positive $\lambda$ only.
The sum includes all energy levels up to some cutoff
$\omega_\mathrm{co}$ at the Thouless 
energy, denoted by the set $\cal{N}$.
It turns out that $H^c$ and $H^s$ do not affect our result, because
they commute with $H^0 + H^p$ and thus leave the eigenstates invariant.
Therefore, it suffices to take $H^p$ -- the BCS pairing Hamiltonian --
as the only interaction term.
Therefore, for our purposes the difference between the
BCS model and 
%the universasl Hamiltonian
the UH 
is only in the
cutoff $\omega_\mathrm{co}$, being  at the Debye energy for the former and at
the Thouless energy for the latter. 
In any case, we define $\DBCS = \omega_\mathrm{co} / \sinh(1/\lambda)$.

The fact that the zero-temperature spectral function
${\cal{A}}_{\hat{O}} (\omega)$  of an
operator $\hat{O}$ is sharply peaked
translates to a strong condition on the matrix elements of the Lehmann representation, 
which is given by
\be
\label{Lehmann}
{\cal{A}}_{\hat{O}} (\omega) = 
\sum_{\ket{I}} \brags \hat{O}^\dagger \ket{I} \bra{I} \hat{O} \ketgs \delta(\omega - E_I).
\ee 
Here $\ketgs$ denotes the ground state, $\ket{I}$ the excited
states with energies $E_I$.
For only one sharp peak to be present in the spectral function, 
the sum in \Eq{Lehmann} must be dominated by one single eigenstate, say 
$\ket{I}\!^0$, whereas all other states $\ket{I} \neq
\ket{I}\!^0$ do not contribute.
Obviously, it will depend on the operator $\hat{O}$ whether this is
the case, and if so, which is the state $\ket{I}\!^0$.
We show that it suffices that $\hat{O}$ satisfies a rather
unrestrictive condition, given after \Eq{A_suitable} below and
fulfilled for many physically relevant quantities.
Furthermore, we show that under this condition, the state
$\ket{I}\!^0$ is from a very limited subset of all possible
excitations, which we characterize below as the ``No-Gaudino states''.
Our finding of well-defined quasiparticles therefore implies that
only these No-Gaudino states are relevant for many physical
properties of systems that satisfy the conditions of the 
%universal Hamiltonian.
UH.

Calculating the spectral function, \Eq{Lehmann}, is usually a
formidable task, equivalent to diagonalizing the Hamiltonian.
Although an exact solution \cite{RichardsonSherman64, DelftRalph01}
exists for the Hamiltonian $H^p$, its complexity in
practice does not allow to calculate spectral functions from it.  
Instead, we use the density-matrix renormalization group (DMRG) method
\cite{WhiteNoack99} for this purpose, a numerical variational approach
that has already been proven very useful for analyzing this model
\cite{DukelskySierra99,SierraDukelsky00,GobertDelft03}.
For suitable operators $\hat{O}$, we are able to obtain the spectral function 
from the DMRG without the usual complications \cite{KuehnerWhite99,Jeckelmann02},
because the state $\ket{I}\!^0$ -- the only one that contributes 
significantly to the spectral function -- can be
constructed explicitly.
The existence of a sum rule allows us to quantify the contribution of other
states  $\ket{I} \neq \ket{I}\!^0$, which we find to be negligibly small.
Finally, we illustrate the use of our method of calculating spectral 
functions by evaluating the tunneling density of states and the 
magnetic response of mesoscopic rings.

\emph{Excitation spectrum and No-Gaudino states:}
Let us begin by describing the excitations of the Hamiltonian
$H^p$ in \eq{pairing_Ham}.
$H^p$ has the well-known property that singly occupied energy
levels do not participate in pair scattering; 
hence their labels (and spins) are good quantum numbers.
Therefore, all eigenstates for which some levels $i$ are singly
occupied are (as far as the remaining levels are
concerned) identical to those of a system with $\cal{N}$ in
\Eq{pairing_Ham} replaced by $\cal{N} \setminus \cal{B}$, where
$\cal{B}$ is the set of singly occupied levels $i$ \cite{DelftRalph01}.
A given state can thus contain two kinds of excitations:
Pair-breaking excitations that go hand in hand with a
change of the quantum numbers $\cal{B}$, and other many-body
excitations that do not.
The latter were studied in \cite{SierraDukelsky03} and dubbed ``Gaudinos''.
In this spirit, we define the No-Gaudino state as the lowest-energy state
within a certain sector of the Hilbert space
characterized by the quantum numbers $\cal{B}$.
%, i.e.~the state that is
%mapped onto the ground state in the presence of the levels 
%$\cal{N} \setminus \cal{B}$.
As is shown below, this state is easily obtained within the DMRG algorithm.

Let us now specify under which condition the spectral function,
\Eq{Lehmann}, is dominated by such a No-Gaudino state.
Any operator can be written as a linear superposition of operators 
\be
\label{A_suitable}
\hat{O} = c^\pdag_{i_1 \sigma_1} \cdots  c^\pdag_{i_k \sigma_k}
c^\dagger_{j_1 \sigma'_1} \cdots c^\dagger_{j_l \sigma'_l}.
\ee
Creating linear superpositions poses no difficulties whatsoever,
therefore it is sufficient to consider operators   of this form.
\emph{The central condition we impose on $\hat{O}$ is that all indices
$i_1, \cdots, j_l$ be mutually different.}
$\hat{O}$ then takes a state with no singly occupied
levels, ${\cal{B}} = \{ \}$, to the sector of the Hilbert space characterized by${\cal{B}} = \{i_1, \cdots, j_l \}$.
We show below that under the above condition, $\hat{O}$ moreover has
the crucial property that when acting on the ground state, it creates
to an excellent approximation the No-Gaudino state in this
sector.
Therefore, the state  $\hat{O}\ketgs$ contributing to the spectral function,
\eq{Lehmann}, is seen to be not only a well-defined eigenstate of the
system, but moreover a No-Gaudino state.

In the BCS limit $d \!\ll\! \DBCS$ (i.e.~at $\lambda \!\gg\! 1/\ln N$,
where $N$ is the number of energy levels within $\omega_\mathrm{co}$),
this follows from the identity 
\begin{eqnarray}
\label{BCS_insight}
\hat{O} \ket{gs} = v_{i_1} \cdots v_{i_k} u_{j_1} \cdots u_{j_l}
\ket{i_1^{-\sigma_1} \cdots  j_l^{\sigma'_l}}^0,
\\
\label{BCS_nogaudino}
\ket{i_1^{-\sigma_1} \cdots  j_l^{\sigma'_l}}^0 =
\gamma^\dagger_{i_1 (-\sigma_1)} \cdots \gamma^\dagger_{j_l \sigma'_l} \ket{gs},
\end{eqnarray}
where the state in \Eq{BCS_nogaudino} is the No-Gaudino state.
Here, $u$, $v$ and $\gamma$ are the coherence factors and the
Bogoliubov quasiparticle operators from BCS theory, as defined e.g.~in
\cite{Tinkham96}.

In the opposite limit $\DBCS \!\ll\! d$ ($\lambda \!\ll\!
1/\ln N$), where perturbation theory in 
$\lambda$ is valid\cite{SchechterDelft01}, the same conclusion is
obtained: to first order 
(i.e. up to errors of order $\lambda^2$), $\hat{O} \ketgs$ 
again creates precisely the No-Gaudino state.
%This statements holds
%as long as the Fermi Level of the intermediate state is not shifted 
%upon application of $\hat{O}$ (clarify!).

There is no such simple analytic argument 
that the Gaudino admixture to $\hat{O}\ketgs$ in \eq{BCS_insight} will
be negligible also in the intermediate regime.
However, this assertion can be checked numerically by a sum
rule, which follows from \eq{Lehmann}:
\be
\label{sumrule}
\int \! {\cal{A}}(\omega) d\omega = 
\sum_{\ket{I}} \bra{gs} \hat{O}^\dagger \ket{I} \bra{I} \hat{O} \ket{gs}
= \bra{gs} \hat{O}^\dagger \hat{O} \ket{gs}.
\ee
%The right hand side is a simple ground state expectation value and is
%therefore easily evaluated using the DMRG. 
We define the lost spectral weight
$w_\mathrm{L} \equiv 
\bra{gs} \hat{O}^\dagger \hat{O} \ket{gs} - 
|\bra{gs} \hat{O}^\dagger \ket{I}^0|^2$
as the part of \eq{sumrule} that is not carried by the No-Gaudino state
$\ket{I}^0$, but instead lost to other background states.
As is shown in \Fig{incl} below, this lost weight turns out to be
negligibly small.

\emph{DMRG algorithm:}
We now give a brief description of the DMRG algorithm as applied to
%the universal Hamiltonian; 
the UH;
more details are described elsewhere
\cite{SierraDukelsky00, GobertDelft03}. 
Energy levels are added one by one to the system
until it obtains its final size.
For simplicity, we assume the energy levels to be equally spaced, although
none of our methods require this assumption.
After adding a level, only a limited number $m$ of basis vectors are kept,
such that the size of the Hilbert space remains numerically manageable.
These basis vectors are selected in order to represent a number of
so-called target states accurately; this is achieved by the DMRG
projection described in \olcite{WhiteNoack99}.
By varying $m$ between $60$ and $140$,
we estimate the relative error in the spectral function from the DMRG
projection to be of the order of $\sim 10^{-5}$ (for $m=60$).
This accuracy 
%is sufficient for our purpose; it 
can be improved by increasing $m$.

In order to calculate the spectral function corresponding to the
operator $\hat{O}$ in \Eq{A_suitable}, we use as target states the
ground state and a state representing the No-Gaudino state 
$\ket{i_1^{-\sigma_1}; \!\cdots ; j_l^{\sigma'_l}}^0$,
%with levels $i_1, \!\cdots, j_l$ singly occupied, 
in the BCS limit given by \eq{BCS_nogaudino}.
%In fact, rather than using the No-Gaudino state itself, we target the
%state
%\be
%\label{DMRG_target}
%\ket{i_1 \!\cdots j_l}^0 \equiv 
%c^\dagger_{i_1 \sigma_1}\!\!\!\cdots c^\dagger_{i_k \sigma_k}
%c^\pdag_{j_1 {\sigma'}_1}\!\!\!\cdots c^\pdag_{j_l \sigma'_l}
%\ket{i_1^{-\sigma_1} \!\!\cdots j_l^{\sigma'_l}}^0, 
%\ee 
%in which these
%levels do, again, not participate in pair scattering, but levels $i_1,
%... i_k$ are now doubly occupied, and levels $j_1, ..., j_l$ are
%empty.  
%We obtain this state by calculating the ground state of a modified
%Hamiltonian, namely \eq{pairing_Ham} with all pair scattering
%involving the levels $i$,$j$ removed. We arrange the levels $i$ ($j$)
%to be occupied (empty) by assigning them a positive (negative) kinetic
%energy term of the order $\pm \oD$.  The main advantage of this choice
%of the target state is that no singly occupied levels occur at any
%point in the algorithm, hence only doubly occupied or empty levels have
%to be considered.  Furthermore, it allows to express 
%Then, the matrix element occurring in the spectral function
%(\ref{Lehmann}) is a simple scalar product: 
%\be
%\label{scalarproduct}
%| ^0\!\bra{i_1^{-\sigma_1} \!\cdots  j_l^{\sigma'_l}}  \hat{O} \ket{gs}| ^2 = 
%| ^0\!\bra{i_1 \!\cdots j_l} gs \rangle |^2.
%\ee
%
%For the excitation energy of the state $\ket{i_1^{-\sigma_1} \!\cdots
%  j_l^{\sigma_l}}^0$, the kinetic energy of the singly occupied levels
%$i_1 \!\cdots j_l$ must be properly accounted for.  In this way, all
%the ingredients for the evaluation of the No-Gaudino contribution to
%\eq{Lehmann} are provided by the DMRG.  
The sum rule, i.e.~the rhs.~of \eq{sumrule}, is evaluated in a separate run with $\ketgs$ and
$\hat{O}^\dagger \ketgs$ as the target states.

%The DMRG results for the spectral function reproduce
%the BCS result, valid for $d \ll \DBCS$, and the result from first
%order perturbation theory in the interaction, valid for $d \gg \DBCS$,
%in the respective limits, but the DMRG is accurate in the entire crossover
%regime as well.

\emph{Dominance of a single No-Gaudino state:}
\begin{figure}
\epsfig{file=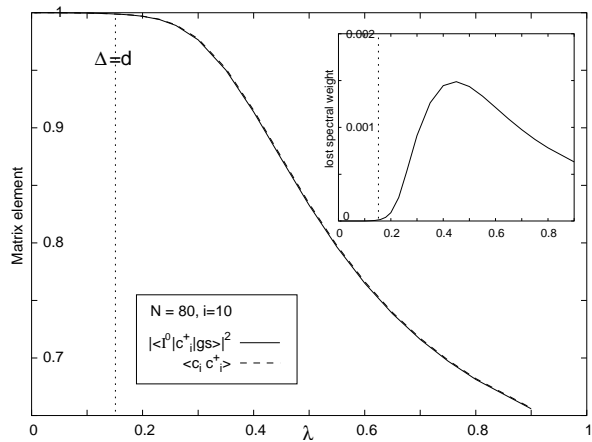, width=.9\linewidth}
 \caption{\label{incl}
The matrix element 
$\langle c_{i\sigma} c^\dagger_{i\sigma} \rangle$ from \Eq{sumrule}
with $\hat{O} = c_{i \sigma}^\dagger$ (dashed 
line) and the contribution from the No-Gaudino state (solid line) as
a function of $\lambda$.
Here, $i=10$ levels above $E_\mathrm{Fermi}$ out of a total
of $N = 2 \cdot 40$ energy levels.
The lost weight $\wL$, i.e. the difference between both, is plotted in
the inset. It shows a maximum in the intermediate regime where 
$d \sim \Delta_\mathrm{BCS}$, but even there, $\wL$ is less than
$0.2\%$ of the total spectral weight.
}
\end{figure}
The fact that the spectral function is dominated by one single
No-Gaudino state is displayed in \Fig{incl}.
Here, the expectation value 
$\brags c_{i \sigma}^{\pdag}  c_{i \sigma}^\dagger \ketgs$, which
occurs in the sum rule, \Eq{sumrule}, with $\hat{O} = c_i^\dagger$,
is plotted (for $i=10$, i.e.~10 levels above $E_\mathrm{Fermi}$)
against the coupling $\lambda$. 
It is practically indistinguishable from the contribution 
$|^0\!\bra{I}  c_{i \sigma}^\dagger \ketgs|^2$ from the
No-Gaudino state only.

The lost weight $\wL$, shown in the inset of \Fig{incl}, 
is seen to be less than  $0.2\%$ of the total spectral weight
throughout the entire parameter regime (for $i=10$; the plots for other
values of $i$, not shown, look similar. The maximum lost weight
somewhat increases as the level $i$ approaches
$E_\mathrm{Fermi}$, but always remains below $1\%$ of the
total weight). 
The lost weight is seen to be vanishingly small for small $\lambda$,
as expected in the perturbative regime $\lambda \!\ll\!1/\ln N$.
Interestingly, the lost weight also decreases for \emph{large} $\lambda$.
This is very untypical for interacting systems, and the underlying reason
is that the dominance of the No-Gaudino state is protected also in the
BCS regime $\lambda \!\gg\!1/\ln N$, see \Eq{BCS_insight}.
% TODO: Quantitative criteria from the gap equation.
Consequently, the lost weight displays a maximum in the crossover
regime at $\lambda \sim 1/\ln N$.
Not shown: We confirmed numerically that the coupling $\lambda_\mathrm{max}(N)$,
at which the lost weight reaches its maximum, always scales linearly
with $1 / \ln N$ as expected. The maximum value 
$w_L(\lambda_\mathrm{max}(N),N)$ turns out to be a monotonically decreasing
function of $N$.

\emph{Applications:}
The dominance of the No-Gaudino state in the spectral function
is not only remarkable by itself, but has also high practical value:
it allows us to calculate the spectral function 
with high precision using the DMRG in
what we call the ``No-Gaudino approximation'' (NGA), in which only the
No-Gaudino state is kept in \Eq{Lehmann}. 
From the spectral function, in turn, many important physical
quantities can be obtained.
The lost weight $\wL$, defined after \Eq{sumrule}, controls the
quality of this approximation: when $\wL$ vanishes, the NGA is exact.

\begin{figure}
\epsfig{file=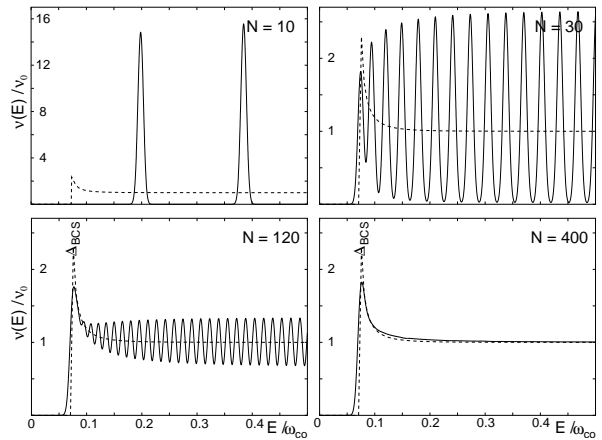, width=.9\linewidth}
 \caption{\label{tunndos_fig}
The tunneling density of states $\nu(E)$ for $\DBCS = 0.07 \oD$ and
$N=10,60,120,400$ energy levels in the No-Gaudino approximation (solid
line; for the sake of better visibility, the delta peaks in
\Eq{Lehmann} have been replaced by Gaussians of width
$0.005 \omega_\mathrm{co}$).
The familiar gap $\DBCS$ emerges during the crossover
from the few-electron ($d\gg\DBCS$) to the bulk limit ($d\ll\DBCS$).
In the latter limit, we observe agreement with the BCS result
(dashed line). 
}
\end{figure}
As a simple first application, we calculate the tunneling density of states 
$\nu(\omega) = \sum_{i\sigma} {\cal{A}}_{c^\dagger_{i\sigma}}(\omega)$
(for $\omega >0$). 
\Fig{tunndos_fig} illustrates that during the crossover
from the few-electron ($d\gg\DBCS$) to the bulk limit ($d\ll\DBCS$),
the familiar BCS gap of width $\DBCS$ emerges together with a strongly pronounced peak
at $\omega \approx \DBCS$ as the quasiparticle energies
are kept away from the Fermi surface by the pairing interaction and
accumulate at $\DBCS$.
The lost weight 
%again shows a maximum at $\lambda \sim 1 / \ln N$, but 
is found to never exceed fractions of $1\%$, thus confirming the
accuracy of the NGA.

\begin{figure}
\epsfig{file=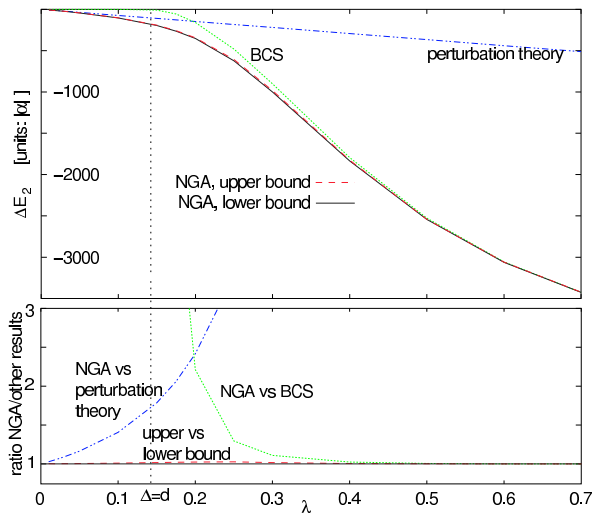, width=.9\linewidth}
 \caption{\label{incl_N40}
Upper part: The magnetic response $\Delta E_2$ from \Eq{E_2} as
function of $\lambda$ for $N=80$ energy levels. The
upper and lower bound from the No-Gaudino approximation (solid and
dashed line) practically coincide and interpolate
between the perturbative and the BCS results (dashed-dotted and dotted
line), valid for small and large $\lambda$, respectively.\\
Lower part: The ratio between the No-Gaudino approximation and the other
results. In the regime $\lambda \sim 1/\ln N$, the perturbative and
the BCS results underestimate the true result (which must lie between
the upper and lower bound) by a factor of more than 2.
}
\end{figure}

As a second example of a quantity that is well captured by
the NGA, we calculate the prediction of the pairing Hamiltonian for the
magnetic response $\Delta E_2$ of small metallic rings
\cite{footnote}, i.e.~the derivative of the persistent
current with respect to flux at zero flux.
%As a second application of the NGA, we calculate the 
%``magnetic response''
%$\Delta E_2$ of small metallic rings, i.e.~the derivative of the
%persistent current with respect to flux at zero flux,
%within the pairing model \footnote{Note that 
%the results from the pairing model, first used in
%\olcite{SchechterOreg02}, 
%differ from those obtained from the local interaction model
%\cite{AmbegaokarEckern90}.
%For a further discussion 
%see \olcite{SchechterLevinson03, EckernAmbegaokar04}.}.
For $\lambda \!\ll\! 1 / \ln N$, $\Delta E_2$ was
calculated in perturbation theory, for $\lambda \!\gg\! 1 / \ln N$ the
BCS approximation was used \cite{SchechterOreg02}.  
Here, using the NGA, we calculate $\Delta E_2$ for all values of
$\lambda > 0$, and specifically in the crossover regime between the
perturbative and the BCS regimes. 

The linear response to the magnetic flux through a ring is given by
$\Delta E_2 = E_2^\mathrm{par} - E_2^\mathrm{dia}$, where
\be
\label{E_2}
E_2^\mathrm{par} \! = \!
-2\left(\frac{e}{m_e L}\right)^2\!\!\!\!\!\sum_{mn, \ket{I}}\!\!\!|P_{mn}|^2 
\frac{ |\bra{I} c^\dagger_{m\uparrow} c^\pdag_{n\uparrow} \!\!-\!  
  c^\dagger_{n\downarrow} c^\pdag_{m\downarrow} \ket{gs}|^2}{E_I}.
\ee
and $E_2^\mathrm{dia}$ equals the $\lambda = 0$ value of
 $E_2^\mathrm{par}$ \cite{SchechterOreg02}.
Here, $m_e$,$e$ are the electron mass and charge, $L$ is the
circumference of the ring.
$P_{mn}$ is the momentum operator between the disordered 1-particle
states, labelled by $m$ and $n$. 
In the highly diffusive regime ($\omega_\mathrm{co} \!<\! 1/\tau$,
where $\tau$ is the
elastic mean free time), which we assume here for simplicity, 
$|P_{mn}|^2 \equiv P^2$ can be taken to be constant for $m \neq n$,
and zero otherwise \olcite{SchechterLevinson03}.
$E_2^\mathrm{par}$ can then be extracted from the spectral functions
for $\hat{O}_{mn\sigma}$ = $c^\dagger_{m\sigma} c^\pdag_{n\sigma}$.

In the NGA, only the states
$\ket{i = m,n; j = n,m}^0$ 
%defined in \Eq{DMRG_target}
are retained in \eq{E_2}.
Because the contribution of the other states, which are
neglected, is always positive, the NGA produces a
lower bound for $E_2$. 
An upper bound can be found as well, namely by
replacing the energy denominator of \eq{E_2} by the energy $E_{mn}^0$of the
No-Gaudino state, which is known to be smaller than the energy of any other
contributing state.
Then, the sum over $\ket{I}$ can be eliminated, 
and the resulting expression for the upper bound is
\be
E_2^> \!=\! 
\alpha \sum_{m \neq n}
\frac{ \bra{gs}
(c^\dagger_{n\uparrow} c^\pdag_{m\uparrow} \!\!-\!
  c^\dagger_{m\downarrow} c^\pdag_{n\downarrow})
( c^\dagger_{m\uparrow} c^\pdag_{n\uparrow} \!\!-\!  
  c^\dagger_{n\downarrow} c^\pdag_{m\downarrow}) \ket{gs}}{E_{mn}^0}.
\ee

The results of our calculation are presented in \Fig{incl_N40}, where
the upper and lower bound is compared to the perturbative and to the
BCS result, given in \cite{SchechterOreg02}. The lower and upper bounds
practically coincide (with an error of $<0.5\%$) in the entire
parameter regime; this reflects the high accuracy of the NGA.  As
expected, the perturbative result is reproduced for 
small $\lambda$ ($\DBCS \ll d$), the BCS result for large $\lambda$
($\DBCS \gg d$).  However, both results underestimate the exact result
by a factor of up to $2.5$ in a large intermediate regime (\Fig{incl_N40} bottom). 
Interestingly, we find that the
magnetic response is much larger than the BCS value also in a regime
in which $\Delta \gg d$, where the BCS approximation is expected to be
valid.
%This is because the contribution of the distant levels from $\DBCS$ up to the
%interaction cutoff, which the BCS approximation does not take into
%account, is important in this regime as well.
This is due to a large contribution of the distant levels from $\DBCS$
up to the interaction cutoff, which the BCS approximation neglects.  A
similarly large contribution from distant levels has previously been
found also for the condensation energy \cite{SchechterDelft01} and
  single particle properties \cite{SchechterLevinson03b}.
%It is known that within the reduced BCS model, the distant levels up
%to the interaction cutoff give a large contribution to physical
%properties such as the condensation energy \cite{SchechterDelft01} and
%single particle properties \cite{SchechterLevinson03b}. 
%Here we show that a similar
%effect occurs in the prediction of this model for
%the magnetic response.

%\emph{Acknowledgments:} 
We thank V.~Ambegaokar, Y.~Imry, D.~Orgad and A.~Schiller for discussions
and ackowledge support through ISF grant No.~193/02-1 and ``Centers of
Excellence'', the DFG program ``semiconductors and metal
clusters'', and the DIP fund. 

\vspace{-0.5cm}
%\bibliography{dfg}

\end{document}